# Surface roughness analysis of the hydrophilic $SiO_2/TiO_2$ nano bi-layers by Level crossing approach

E. Daryaei [1], M. Reza Rahimi Tabar [1, 2] and A. Z. Moshfegh [1, 3] [1]

[1] Department of Physics, Sharif University of Technology, PO Box 11155-9161, Tehran, Iran

[2] Institute of Physics, Carl von Ossietzky University, 26111 Oldenburg, Germany

[3] Institute for Nanoscience and Nanotechnology, Sharif University of Technology, PO Box 14588-89694, Tehran, Iran

**Abstract**

The effect of etching time on the statistical properties of the hydrophilic surface of $SiO2/TiO_2$/Glass nano bi-layer has been studied using Atomic Force Microscopy (AFM) and stochastic approach based on the level crossing analysis. We have created a rough surface of the hydrophilic $SiO_2/TiO_2$ nano bi-layer system by using 26% Potassium Hydroxide (KOH) solution. Measuring the average apparent contact angle assessed the degree of hydrophilicity and the optimum condition was determined at 10 min etching time. Level crossing analysis based on AF images provided deeper insight into the microscopic details of the surface topography. For different etching time, it has been shown that the average frequency of visiting a height with positive slope behaves Gaussian for heights near the mean value and obeys power law for the heights far away from the mean value. Finally, by applying the generalized total number of crossings with positive slope, it was found that the both high heights and deep valleys of the surface are extremely effective in hydrophilic degree of the $SiO_2/TiO_2$/Glass nano bi-layer investigated system.

[1] Corresponding author
moshfegh@sharif.edu
Fax: +98- 21- 6601- 2483
Tel: +98- 21- 6616- 4516

# 1. Introduction

Whenever a droplet of liquid such as water is placed on the surface of a substrate, it tries to reach an equilibrium state and the surface becomes wet [1]. The degree of wetting is determined by balance between adhesive and cohesive forces. When water spreads and covers the surface in the macroscopic scales, it is called the surface is hydrophilic. The ability to control the degree of hydrophilicity of a solid surface is extremely important and useful in different range of technological applications. In the last decade, extensive efforts have been focused on application of hydrophilic surfaces such as: self-cleaning surfaces, anti-fogging mirrors [2] and photocatalysts [3,4].

It has been established that the degree of the wetting and spreading generally depend on both external conditions such as temperature [1] and internal conditions like surface properties. Considering the later, it has been shown that the surface topography of a substrate [5,6,7] and surface impurities and contamination as chemical parameter have the most effective role in hydrophilic degree (or apparent contact angle) of a given hydrophilic surface [8,9,10]

Today, surface roughness has a vast consideration in science and technology [11,12], it has an important effect in some physical phenomena such as: friction, degrees of hydrophilicity and hydrophobicity, self-cleaning, [13,14] and also improving the mass throughput in microchannel and nanochannel flows [15].

Effect of surface roughness on contact angle (and hence hydrophilicity) was studied by many researchers [16]. Norman Morrow carried out pioneering studies of the effect of surface roughness on contact angle, and reported an excellent and extensive set of data [17]. Others studied the effect of surface roughness on moving contact angles [18-20].

Recently, superhydrophilicity of $SiO_2/TiO_2$ thin film system has been reported by our group and other researchers [21-22-23]. It was reported that the hydrophilicity of the $SiO_2/TiO_2$ nano bi-layer films is due to the stable Si–OH groups and the photo-catalytic $TiO_2$ under-layer maintains the hydrophilicity of the double layer films by decomposing organic contaminants on the film surface [22]. Moreover, $SiO_2/TiO_2$ nano bi-layer films exhibit a natural, persistent and regenerable superhydrophilicity without the need of UV light [24].

Here, first, SiO$_2$/TiO$_2$/Glass nano bi-layer samples have been prepared by RF sputtering technique, then, by using 26% Potassium Hydroxide (KOH) solution, surface morphology and surface roughness of the layers were examined under different etching time. Then their degrees of hydrophilicity have been measured after three weeks.

To understand a deeper insight into the variation of surface topography during the etching process of the hydrophilic SiO$_2$/TiO$_2$/Glass nano bi-layer system, we have applied level crossing analysis on images obtained by atomic force microscopy (AFM). The level crossing analysis of this data has the advantage that it provides important global properties of a surface. A detailed foundation and principle of level crossing analysis approach for rough surfaces has been described in [25]. This stochastic tool has been used to measure the surface roughness for different systems such as Co(3 nm)/NiO(30 nm)/Si(100) structure used in the magnetic multilayers [26], effect of annealing temperature on the statistical properties of WO$_3$ surface using atomic force microscopy (AFM) [27] and laser-induced silicon surface modification [28]. In addition, this method was also employed in studying the fluctuations of other systems with scale dependent complexity, such as: fluctuations of velocity fields in Burgers turbulence [29], Kardar–Parisi–Zhang equation in (d + 1)-dimensions [30], stock market [31]. As a new application, we have used the level crossing analysis to describe the effect of etching time on the hydrophilic degrees of SiO$_2$/TiO$_2$ nano bi-layer system.

## 2. Level Crossing Analysis: Theoretical Approach

Level crossing analysis of stochastic process has been introduced first by Rice in 1944 [32]. This method was applied to investigate statistical properties of rough surfaces [25, 26]. Consider a surface with size L×L which has been grown (or etched) in laboratory, so each point on surface such as $x$ has a height $h(x)$, this height function behaves stochastic because the etching process is nondeterministic. Thus, we can apply level crossing analysis to investigate some general properties of an etched surface.

In the level crossing analysis, we are interested in determining average number of visiting the definite value for a stochastic variable such as the height function $h(x) - \bar{h} = \alpha$ at a surface where $\bar{h}$ is the ensemble average of a stochastic height function, $h(x)$, with positive slope in a sample with size L, $N_\alpha^+$ (see Figure 1) [25]. For a statistically homogeneous process, the average number of crossings is proportional to the space interval L, hence $N_\alpha^+$ is proportional to L or $N_\alpha^+ = \nu_\alpha^+ L$ where $\nu_\alpha^+$ is the average frequency (in spatial dimension) of observing the given height α with positive slope. This function contains almost all statistical information present on the surface and shows how the stochastic height $h(x)$ behaves. The frequency $\nu_\alpha^+$ can be written in terms of a joint Probability Distribution Function (PDF) defined by $P(\alpha, \dot{h})$ as follows [25]:

$$\nu_\alpha^+ = \int_0^\infty P(\alpha, \dot{h}) \dot{h} d\dot{h} \qquad (1)$$

Where $\dot{h} = (h(x+\Delta x) - h(x))/\Delta x$ and $\Delta x$ is the differential length scale. Therefore, we can count the number of visiting a definite height such as α and obtain general properties of joint PDF function and all statistical details for a surface. The magnitude of $\nu_\alpha^+$ can be related to magnitude of 3D surface area, so its determination is one of the most suitable approaches in experimental surface analysis [26].

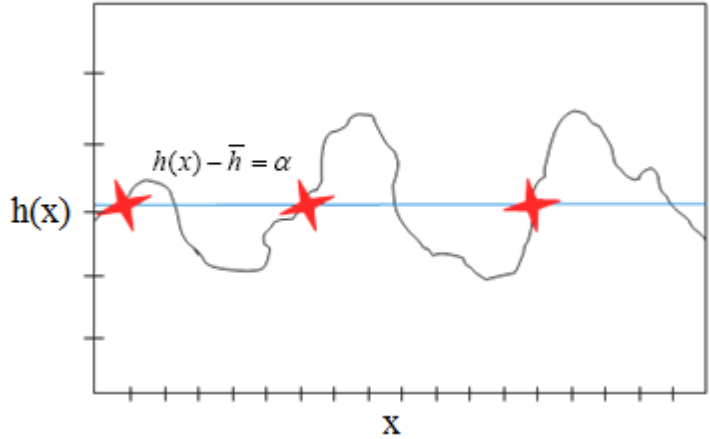

**Figure 1**. Schematic positive slope crossing in a fixed α level in a rough surface.

As introduced in Ref [25-31], we can also define the generalized total number of crossings with positive slope, $N_{tot}^+(q)$ as described by

$$N_{tot}^+(q) = \int_{-\infty}^{+\infty} \nu_\alpha^+ |\alpha|^q d\alpha \qquad (2)$$

where zero moment $q = 0$ (with respect to $v_\alpha^+$) shows the total number of crossings for a height with positive slope $N_{tot}^+ = \int_{-\infty}^{\infty} d\alpha v_\alpha^+$, The moments $q < 1$ will give information about the frequent events while moments $q > 1$ are sensitive to the rare events. The moment $q=1$, will measure the total number of crossings of a surface with positive slope multiplied by their heights. So, the $N_{tot}^+(q=1)$ can measure the captive total volume and square area of the surface are in the same order [26].

## 3. Experimental details

The $2 \times 1 cm^2$ glass substrates were cleaned in an ultrasonic bath with high purity methanol for 10 minutes and the residual impurities cleaned with distilled water for a few times to assure surface cleanness. The $SiO_2$(20 nm)/$TiO_2$(80 nm) nano bi-layers were grown on the cleaned glass substrate by using RF sputtering deposition technique at room temperature (RT). Two different Ti and Si targets were used under a mixed Ar (60%) and $O_2$ (40%) gas discharge at total pressure of 10 mTorr.

The deposited $SiO_2$ (20 nm)/$TiO_2$ (80 nm) nano bi-layers were subsequently etched by applying 26% solution of Potassium Hydroxide (KOH). For preparing KOH solution, 80ml DI water, 25ml propanol (>99.9%, Merck) and 26gr KOH (>99.9%, Merck) mixed together. The KOH etching process was performed on the prepared bi-layers in a constant temperature bath at 71 $^0$C using clean quartz glassware in various etching times namely 4, 8, 10 and 12min to obtained optimum etching time as compared with unetched surface (zero time).

After three weeks of KOH etched $SiO_2$ surface of the $SiO_2$/$TiO_2$/Glass nano bi-layer system where the samples had been in normal atmospheric conditions (relative humidity and temperature were about 60% and 25 $^0$C, respectively), the water apparent contact angle measurements were performed in atmospheric air at room temperature by employing a commercial contact angle meter (Dataphysics OCA 15 plus) with $\pm 1^0$ accuracy. To do that, a water droplet was injected on several spots of the surface using a $2\mu l$ micro-injector for obtaining accurate statistical analysis.

In order to determine the apparent contact angle of the clean etched surface (without any surface dirtiness), the etching process was repeated for the optimum etching time and its apparent contact angle was measured again. Surface roughness (the standard deviation of the height values within the given area of AFM image) and surface topography of the samples were characterized by Thermo Microscope Autoprobe CP-Research atomic force microscopy (AFM) in air with a silicon tip of 10 nm radius, in contact mode and the images digitized into 1024×1024 pixels.

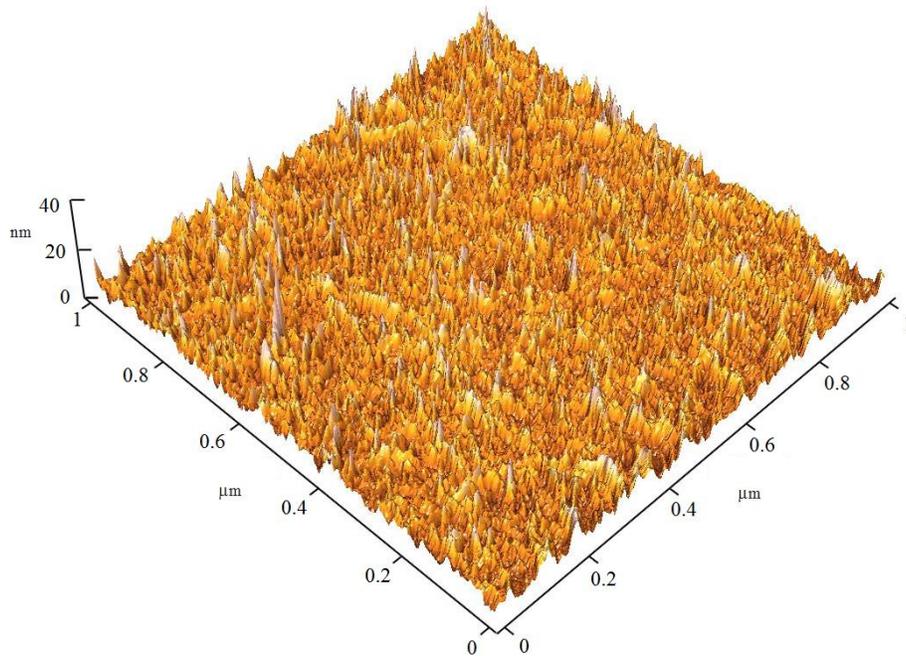

**Figure 2.** Three-dimensional AFM image of the $SiO_2$/$TiO_2$/Glass nano bi-layer surface observed after 10 min etching time

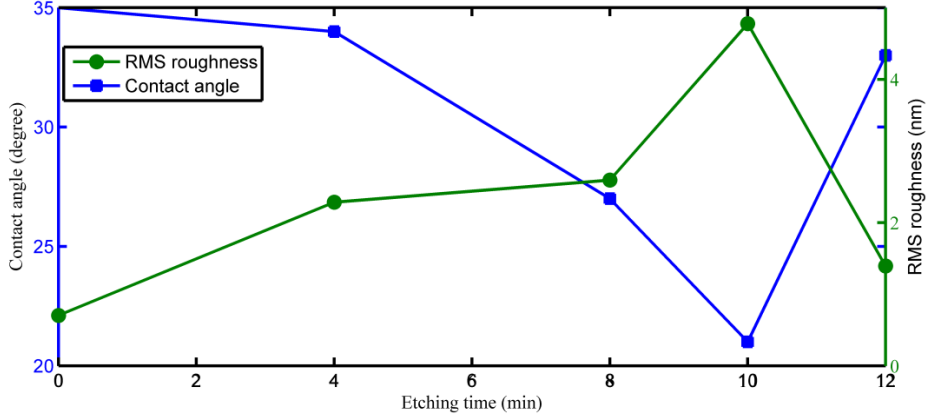

**Figure 3.** Variation of average contact angle and the RMS surface roughness of the $SiO_2/TiO_2/Glass$ nano bi-layer surface as function of etching time (all experimental data obtained after three weeks kept in atmospheric condition).

## 4. Results and discussion

According to our experimental results, the lowest apparent contact angle obtained at ~ $4^0$ for the clean etched surface just after 10 minutes etching time. Nevertheless, it increased to $21^0$ after three weeks duration. The observed increasing trend is consistent with previous study [33]. The behavior of this system is associated with amount of surface OH species determined by X-ray photoelectron spectroscopy (XPS) reported by our group [23] and others (see for example [34]). Surface topography of the all etched $SiO_2/TiO_2/Glass$ nano bi-layer system was also investigated by AFM technique. The three-dimensional AFM image is illustrated in Figure 2 indicates average nanostructure size and surface roughness after 10 minute etching time.

Figure 3 shows both changes in the average water apparent contact angle on the surface and changes in the RMS surface roughness of the $SiO_2/TiO_2/Glass$ nano bi-layer in various etching times. The RMS roughness curve indicates that surface roughness increased from 0.7 nm to 4.8 nm with increasing etching time. Nevertheless, it decreased again to 1.4 nm after 12 minute etching time.

It is necessary to note that all contact angle measurements shown in Figure 3 were performed three weeks after etching of the $SiO_2$ surface to assure surface dirtiness, without any surface cleaning process such as $UV/O_3$ treatment [33]. The minimum apparent contact angle was determined at $21^0$ after 10 min of etching time. In general, by increasing the etching time, the apparent contact angle decreases and after 10 min it begins to increase due to reduction of $SiO_2$ surface roughness. This reduction is originated from decreasing thickness of $SiO_2$ layer with increasing etching time. The improvement of the degree of hydrophilicity of the $SiO_2/TiO_2$/Glass nano bi-layer system by the etching process is the result of topographical change of etched surface. Increased surface area and topographical changes cause a reduction in the apparent contact angle measurements. This improvement was also observed in other etched substrates [35,36,37].

There are two conventional theories to discuss the effect of surface roughness on the degree of the hydrophilicity: Wenzel's theory [5] and Cassie-Baxter's theory [6]. The former describes a homogeneous wetting regime that the droplet completely covers the rough surface without any air packet while the later is based on the idea that some entrapped air in a rough heterogeneous surface could enhance its hydrophobicity recognizing that the water drop is partially sitting on air, not filling the valley. Wenzel equation defined as $\cos(\theta_W) = r\cos(\theta_Y)$, $\cos(\theta_W) = r\cos(\theta_Y)$, where $\theta_W$, $\theta_Y$ and $r$ are Wenzel's contact angle for rough surface, Young contact angle for a smooth surface and Wenzel's roughness factor (which is defined as the ratio of the actual area of rough surface to the geometric projected area), respectively. In a more general and real surface, it can be described by Cassie-Baxter equation defined as $\cos(\theta_{CB}) = r\,f\cos(\theta_Y) + f - 1$, where the $\theta_{CB}$ and $f$ are Cassie-Baxter's contact angle for heterogeneous surface and the fraction of solid surface area wet by the liquid, respectively.

In this study, based on our experimental data analysis, we have used Wenzel model to describe degree of the hydrophobicity of the etched surface. This was done by calculating $\cos^{-1}(\cos\theta_W / r)$ for all etched surfaces in order to determine apparent contact angle for smooth surface (removing surface roughness effect). We have obtained the values of $r$ factor from AFM data analysis. As a result, contact angle for smooth surface was found in the range from $41^0$ to $47^0$ for $r$ value from 1.20 to 1.32, for all investigated samples. These findings are in good agreement with Wenzel's regime. We believe that due to surface dirtiness of the $SiO_2/TiO_2$/Glass nano bi-layer, we can ignore the hydrophilic role of $TiO_2$ sub-layer. It is important to note that the obtained $r$ factor is related to both AFM image scale and resolution of each image. We estimated the $r$ factor from the AFM images with size $1\times1\mu m^2$ and $1024\times1024$ pixels.

According to water apparent contact angle measurements, etching rate of the used 26% KOH solution at 70 $^0$C was determined 15 Angstrom per minute by observing unchanged apparent contact angle after 13 min approximately. This is related to complete etching of the SiO$_2$ layer (the TiO$_2$ layer cannot be etched by the KOH solution).

To find a detailed knowledge on the variation of surface topography during the etching process of the hydrophilic SiO$_2$/TiO$_2$/Glass nano bi-layer system, we have applied level crossing analysis on images obtained by atomic force microscopy (AFM). This procedure was performed on the scanned surfaces as two dimensional stochastic data as following: first, the surfaces were cut at different height levels i.e. $h(x) - \bar{h} = \alpha$ [25] in x and y directions, and then, the number of crossing points with positive slope at each level (as shown in figure 1) were counted to determine a particular frequency $\nu_\alpha^+$.

Figure 4 shows the average frequency of visiting the height α with positive slope, $\nu_\alpha^+$, as a function of α in different etching time. It is seen that by increasing etching time, the average frequency of visiting the height α ($\nu_\alpha^+$) becomes more broadening until 10 minute and after that it narrows down. In fact, the height dependence of the $\nu_\alpha^+$ function is a Gaussian for small α`s that related to the frequently visited heights near the average height for all surfaces. The standard deviation of each etching time curve (Fig.4) is related to corresponding surface roughness obtained by the conventional AFM analysis as seen in Figure 3. The shifted peak position of the distribution must be related to the tip convolution effect that was also reported in our previous studies [26,27].

However, there is a definite departure from the Gaussian distribution at large α for all etched samples. The average frequency of visiting the height for height far from the mean value where are related to rare events, behaves as power law i.e. $\nu_\alpha^+ \sim \alpha^{-b}$, where $b$ exponent is related to surface topography and changes for different etching time. The exponent was obtained by using a least-squares fitting method for different etching time (see the inset of Figure 4). It is varying from 2.2 to 8.5 for respectively 10 and 12 min etched samples. The small exponent originates from the increase in number of high heights and deep valleys present on the etched surface.

In fact, the behavior of several physical, at least in some range of length or time scales, is dominated by large and rare fluctuations that are characterized by broad distributions with power-law tails [38]. However, there are several reports of power law behavior in some nanostructural systems by using AFM data such as edge distribution of nanowire in inorganic conducting network [39] but, to the best of our knowledge, this is the first report on observation of power law distribution for heights on the hydrophilic surface of $SiO_2/TiO_2$/Glass nano bi-layer using AFM images.

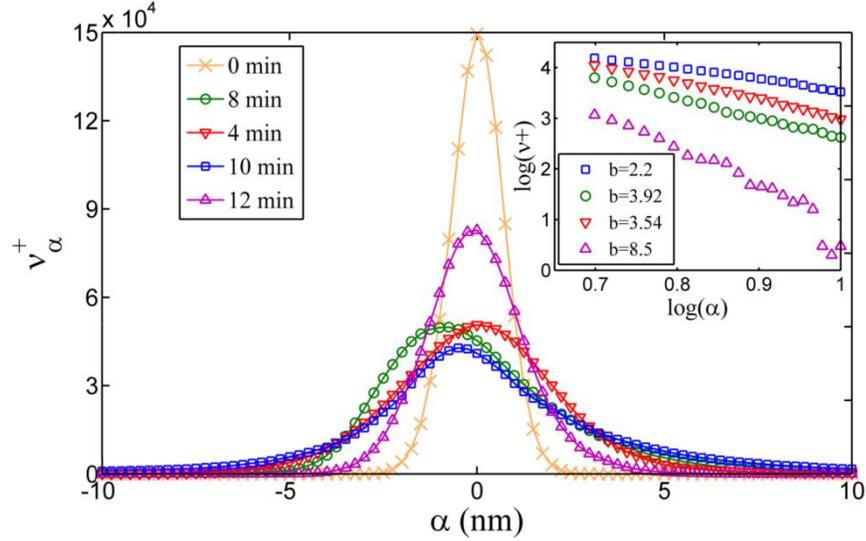

Figure 4. The average frequency of visiting height $h(x) - \bar{h} = \alpha$ with positive slope $v_\alpha^+$ of the etched $SiO_2/TiO_2$/Glass nano bi-layer at different etching times as a function of α. Inset: power law behavior for heights far from mean value (for large α)

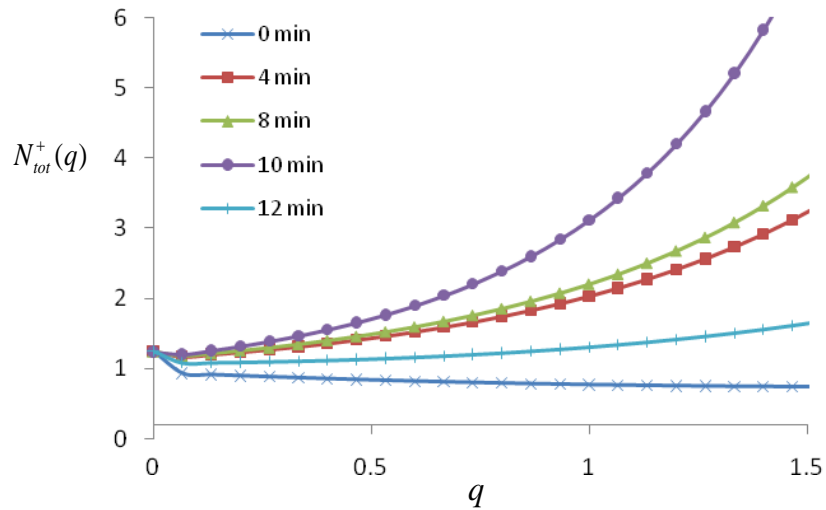

Figure 5. Generalized total number of crossings with positive slope $N_{tot}^{+}(q)$ for the etched SiO$_2$/TiO$_2$/Glass nano bi-layer at different etching times.

In order to get more information about different moment of the probability distribution function, one could calculate the generalized total number of crossings with positive slope, i.e. equation (2). For the moment with $q = 0$, it shows the total number of crossings for a height with positive slope. Indeed it shows the total number of an AFM image that is equal to number for all pixels, i.e. $1024 \times 1024 \sim 10^6$. For moment $q = 1$, will be proportional to the total area of 3D surface of nanostructure of the etched SiO$_2$ surface as reported for sputtered Co (3 nm)/NiO(30 nm)/Si(100) structure [21]. The higher moments, $q > 1$, give information about the tail of PDF for a rough surface. .For q>>1 the high values of $\alpha$ will dominant in the Eq. 2 . Using the power law expression for level crossing frequency one finds:

$$N_{tot}^{+}(q >> 1) \sim d^{q-b}$$

where $d$ is proportional to the SiO$_2$ layer thickness as a limit of integration for Eq. 2.

Figure 5 shows the generalized total number of crossings with positive slope, $N_{tot}^{+}(q)$ for the etched SiO$_2$/TiO$_2$/Glass nano bi-layer surface at different etching times. This function has an increasing trend by increasing the $q$ value in all etched samples. Base on this figure, it was found that the sample with high degree of the hydrophilicity had a long tail in PDF function. This means that the number of high heights and deep valleys are much larger than other etched surfaces. Therefore, the role of extreme events in the height fluctuations is effective in improving of the hydrophilicity of the SiO$_2$/TiO$_2$/Glass surface.

For some applications in which higher surface area are useful (such as hydrophilic surfaces of photocatalysts at Wenzel's regime), considering the behavior of $N_{tot}^{+}(q=1)$ is more appropriate. The increase in $N_{tot}^{+}(q=1)$ has the same meaning as the increase in both surface area and Wenzel's roughness factor. Indeed, for two surfaces with the same height standard deviation, $N_{tot}^{+}(q=1)$ can separate the one with a higher surface area.

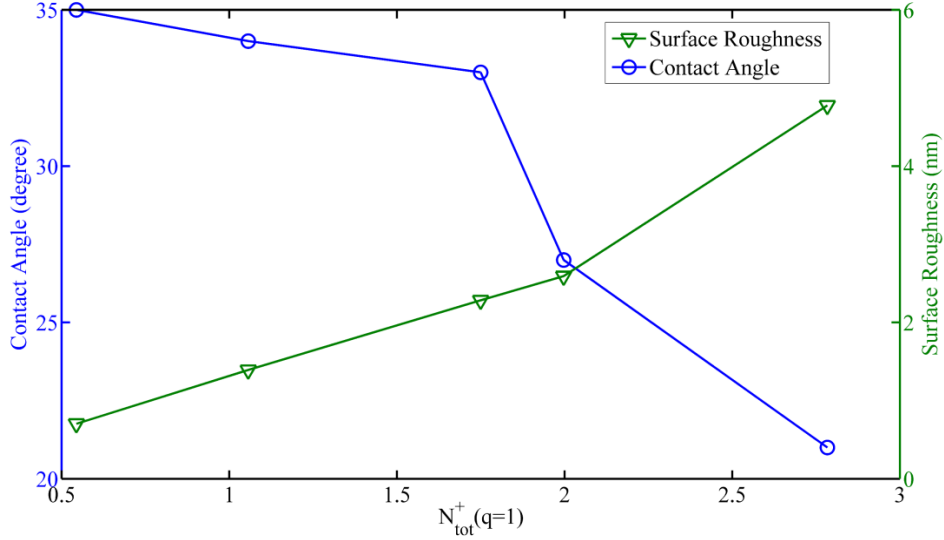

Figure 6: Variation of surface roughness and contact angle via the generalized function $N_{tot}^{+}(q=1)$

To correlate the measured macroscopic property (apparent contact angle) with microscopic property (surface roughness) of the etched surfaces, we have applied $N_{tot}^{+}(q=1)$ function as a tool to measure and characterize the three-dimensional etched surface. Figure 6 shows the variation of surface roughness and contact angle of etched surface via generalized total number of crossing with positive slope at $q=1$. It is clear that the contact angle is reduced with increasing $N_{tot}^{+}(q=1)$, whereas the surface roughness of the $SiO_2$ etched layer has an increasing trend but not in a linear fashion. Thus, we can consider $N_{tot}^{+}(q=1)$ as another quantity to evaluate surface morphology of the samples. It is necessary to note that in this figure, all the points in the both graphs are arranged based on increasing $N_{tot}^{+}(q=1)$ value are not arranged by the etching time duration.

## 5. Conclusions

The nanoscale roughness was generated on hydrophilic surface of sputtered $SiO_2$/$TiO_2$/Glass nano bi-layers by applying 26% KOH solution. The results of apparent contact angle measurement and AFM analysis have shown that extreme events in nanoscale surface roughness significantly increase the degree of the surface hydrophilicity. These microscopic changes in surface topography contribute in increasing the hydrophilicity. A behavioral change from increasing hydrophilicity to decreasing hydrophilicity observed after 10 minutes of etching time. We have found that the hydrophilicity degree of a surface can be improved by using a simple etching process.

To obtain a deeper insight into the variation of surface topography during the etching process of the hydrophilic $SiO_2/TiO_2/Glass$ nano bi-layers system, we have applied level crossing analysis on images obtained by AFM. We have investigated the role of etching time, as an external parameter, to control the statistical properties of a rough $SiO_2$ surface. Moreover, by using the level crossing analysis, we have obtained an optimum etching time (10 min) to have the highest hydrophilic degree. The average frequency of visiting the height α with positive slope, $v_\alpha^+$ was calculated for different etching time and it was found that the height dependence of the $v_\alpha^+$ is Gaussian for heights near the mean value and it obeys power law for heights far from the mean value. Applying the generalized total number of crossings with positive slope, $N_{tot}^+(q)$ we have found that the high heights and deep valleys of the surface are extremely effective in hydrophilic degree of the $SiO_2/TiO_2/Glass$ nano bi-layer.

## Acknowledgments

We would like to thank Dr. G.R. Jafari, Dr. R. Azimirad and Dr. O. Akhavan for useful discussion, and F. Vaseghiniya for providing AFM images. A. Z. M. acknowledges the financial support from Sharif University of Technology.

# References

[1] For a recent review, see Bonn D, Eggers J, Indekeu J, Meunier J and Rolley E 2009 *Rev.Mod. Phys.* **81** 739

[2] Fujishima A, Hashimoto K, Watanabe T 1999 *TiO2, photocatalysis, fundamentals and applications* (BKC Inc. Tokyo Japan)

[3] Fujishima A, Zhang X and Tryk D A 2008 *Surf. Sci. Rep.* **63** 515

[4] Moshfegh A Z 2009 *J. Phys. D: Appl. Phys.* **42** 233001

[5] Wenzel R.N 1936 *Ind. Eng. Chem.* **28** 988 ; Wenzel R N 1949 *J. Phys. Colloid Chem.* **53** 1466

[6] Cassie A.B.D. and Baxter S. 1944 *Trans. Faraday Soc.* **40** 546

[7] Quéré D 2005 *Rep. Prog. Phys.* **68** 2495

[8] Okada K, Tomita T, Kameshima Y, Yasumori A and MacKenzie K J D 1999 *J. Colloid Interface Sci.* **219** 195

[9 ] Soeno T, Inokuchi K and Shiratori S 2004 *Appl. Surf. Sci,* **237** 539

[10] Azimirad R, Naseri N, Akhavan O and Moshfegh A. Z 2007 *J. Phys. D: Appl. Phys.* **40** 1134

[11] Barabási A.L and Stanley H. E 1995 *Fractal concepts in surface growth* (Cambridge University Press); G. R. Jafari, S. M. Fazeli, F. Ghasemi, S. M. Vaez Allaei, M. Reza Rahimi Tabar, A. Iraji zad and G. Kavei 2003 Phys. Rev. Lett. **91,** 226101

[12] Davies S and Hall P 1999 *J. Roy. Stat. Soc.* B **61** 3

[13] Peressadko A. G, Hosoda N and Persson B. N. J. 2005 *Phys. Rev .Lett.* **95** 124301

[14] Zhao Y. P, Wang L. S. and Yu T. X. 2003 *J. Adhes. Sci. Technol.* **17** 519

[15] Sbragaglia M, Benzi R, Biferale L, Succi S and F. Toschi 2006 *Phys. Rev .Lett.* **97** 204503

[16] Sahimi M., *Flow and Transport in Porous Media and Fractured Rock*, 2nd ed., Chap. 14 (Wiley-VCH, Weinheim, 2011)

[17] Morrow, N.R, Ind. Eng. Chem. 62 , 32 (1970); Morrow, N.R., J. Can. Pet. Technol. 14 , 42 (1975); Morrow, N.R., J. Can. Pet. Tech. 15 , 49 (1976)

[18] Joanny, J.F., and P.G. de Gennes, J. Chem. Phys. 81 , 552 (1984)

[19] Pomeau, Y., and J. Vannimenus, J. Colloid Interface Sci. 104 , 447 (1985)

[20] Joanny, J.F., and M.O. Robbins, J. Chem. Phys. 92, 3206 (1990)

[21 ] Nakamura M 2006 Thin Solid Films 496 131

[22] Nakamura M, Kobayashi M, Kuzuya N, Komatsu T and Mochizuka T, 2006 Thin Solid Films 502 121

[23] Ganjoo S, Azimirad R, Akhavan O and Moshfegh A Z 2009 J. Phys. D: Appl. Phys. 42 025302

[24] Permpoon S, Houmard M, Riassetto D, Rapenne L, Berthomé G, Baroux B, Joud J.C. and Langlet M 2008 Thin Solid Films, 516 957

[25] Friedrich R, Peinke J, Sahimi M and Reza Rahimi Tabar M 2011 Physics Reports 506, 87; Shahbazi F, Sobhanian S, Reza Rahimi Tabar M, Khorram S, Frootan G R and Zahed H, 2003 J. Phys. A Math. Gen. 36 2517

[26] Sangpour P, Jafari G R, Akhavan O, Moshfegh A Z and Reza Rahimi Tabar M 2005 Phys. Rev. B 71 155423

[27] Jafari G R, Saberi A A, Azimirad R, Moshfegh A Z and Rouhani 2006 S J. Stat. Mech. P09017

[28] Vahabi M Jafari G R, Mansour N, Karimzadeh R and Zamiranvari J 2008 J. Stat. Mech. P03002

[29] Movahed S, Bahraminasab A, Rezazadeh H and Masoudi A A 2006 J. Phys. A: Math. Gen 39 3903

[30] Bahraminasab A, Sadegh Movahed M, Nassiri S D and Masoudi A A 2005 Preprint condmat/0508180

[31] Jafari G R, Movahed M S, Fazeli S M, Reza Rahimi Tabar M and Masoudi S F, 2006 J. Stat. Mech. P06008; F. Shayeganfar, M. Hölling, J. Peinke, and M. Reza Rahimi Tabar, 2012, J. Physica A391, 209.

[32] S. O. Rice 1944 Bell System Tech. J. 23 282

[33] Takeda S and Fukawa M 2005 Mater. Sci. Eng. B 119 265

[34] Takeda S and Fukawa M 2003 Thin Solid Films 444 153

[35] Tahk D, Kim T, Yoon H, Choi M, Shin K and Suh K. Y 2010 Langmuir, 26, 2240

[36] Zenglin W, Zhixin L, Yue H and Zhixiang W, 2011 Journal of The Electrochemical Society, 158 (11) D664

[37] Dongseob K, Sangmin L and Woonbong H 2012 Current Applied Physics 12 219 15

[38] Amaral L.A.N, Buldyrev S.V, Havlin S, Salinger M.A and Stanley H.E 1998 Phys. Rev. Lett. 80, 1385; Viswanathan G.M., Buldyrev S.V, Havlin S, da Luz M. G. E, Raposo E. P and Stanley H. E 1999 Nature (London) 401, 911;Barabasi A. L. and Albert R. 1999 Science 286, 509

[39] Strle J, Vengust D and Mihailovic D 2009 Nano Lett. 9 1091